\documentclass[twocolumn,10pt,amsmath,amssymb,superscriptaddress,aps,prb,longbibliography]{revtex4-2}
\usepackage[dvipdfmx]{graphicx}
\usepackage{newtx}
\usepackage{physics}
\usepackage[version=4]{mhchem}
\usepackage{microtype}
\usepackage{siunitx}
\usepackage{miller}
\usepackage{color}

\begin{document}

\title{Closing of the Mott gap near step edges in \ce{NiS2}}

\author{Yuuki~Yasui}
\email{yasui@k.u-tokyo.ac.jp}
\affiliation{RIKEN Center for Emergent Matter Science, Wako, Saitama 351-0198, Japan}
\affiliation{Department of Advanced Materials Science, The University of Tokyo, Kashiwa, Chiba 277-8561, Japan}

\author{Kota~Iwata}
\affiliation{Department of Advanced Materials Science, The University of Tokyo, Kashiwa, Chiba 277-8561, Japan}

\author{Shota~Okazaki}
\affiliation{Laboratory for Materials and Structures, Tokyo Institute of Technology, Yokohama, Kanagawa, 226-8501, Japan}

\author{Shigeki~Miyasaka}
\affiliation{Department of Physics, Osaka University, Toyonaka, Osaka, 560-0043, Japan}

\author{Yoshiaki~Sugimoto}
\affiliation{Department of Advanced Materials Science, The University of Tokyo, Kashiwa, Chiba 277-8561, Japan}

\author{Tetsuo~Hanaguri}
\email{hanaguri@riken.jp}
\affiliation{RIKEN Center for Emergent Matter Science, Wako, Saitama 351-0198, Japan}

\author{Hidenori~Takagi}
\affiliation{Department of Physics, Graduate School of Science, The University of Tokyo, Bunkyo, Tokyo 113-0033, Japan}
\affiliation{Institute for Functional Matter and Quantum Technologies, University of Stuttgart, 70569 Stuttgart, Germany}
\affiliation{Max Planck Institute for Solid State Research, Heisenbergstra\ss e 1, 70569 Stuttgart, Germany}

\author{Takao~Sasagawa}
\affiliation{Laboratory for Materials and Structures, Tokyo Institute of Technology, Yokohama, Kanagawa, 226-8501, Japan}
\affiliation{Research Center for Autonomous Systems Materialogy, Tokyo Institute of Technology, Yokohama, Kanagawa 226-8501, Japan}

\date{\today}

\begin{abstract}
	A prototypical charge-transfer type Mott insulator \ce{NiS2} pyrite exhibits a metal-insulator transition with bandwidth control.
	Recent discoveries on surface-specific electronic states on other $3d$ transition-metal disulfide pyrites motivate us to further investigate the surface of \ce{NiS2}, where metallic surface conduction is discussed.
	Here, the spectroscopic-imaging scanning-tunneling-microscopy observations revealed that the surface is not metallic, contrary to the expectation.
	Instead, the Mott gap is closed near step edges, suggesting possible electrical conduction from one-dimensional channels.
	The edge anomaly was observed irrespective of its magnetic order and is limited to the insulator phases.
\end{abstract}

\maketitle

\begin{figure}[b]
	\includegraphics[width=\columnwidth]{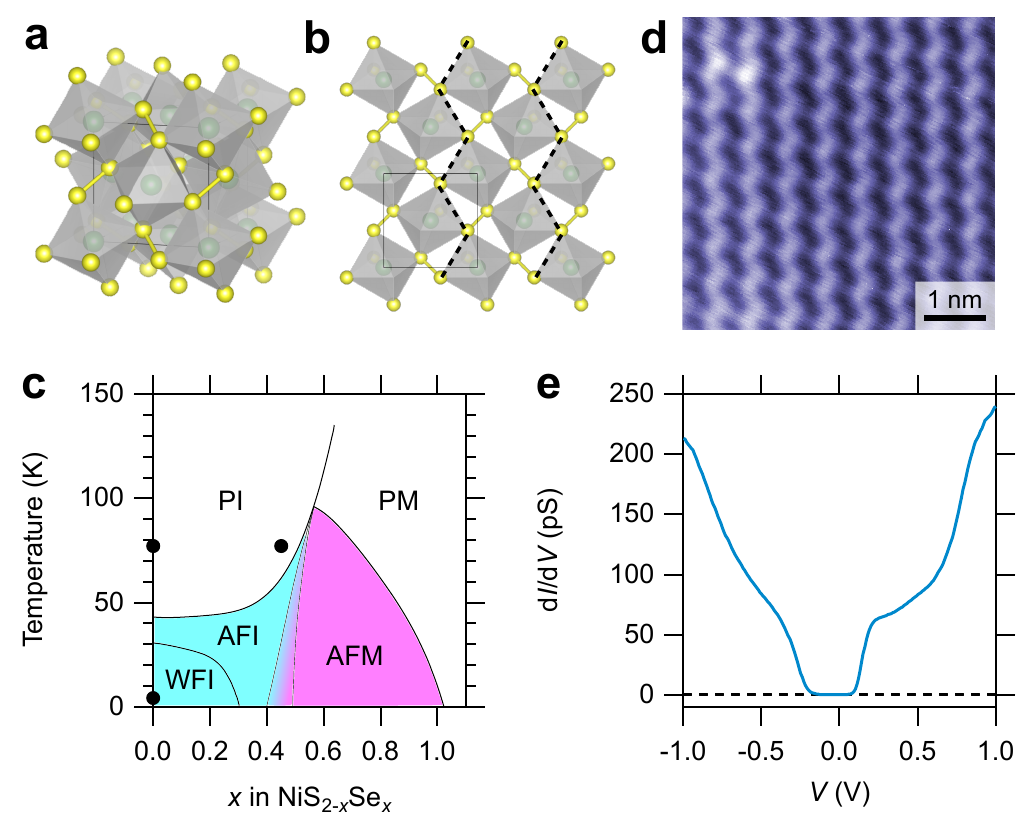}
	\caption{
	(a)~The crystal structure of \ce{NiS2}~\cite{momma_2011_jac, jain_2013_aplm}.
	Ni atoms are represented in green, and S atoms are in yellow.
	(b)~A model for a cleaved surface seen from the \textit{c}-axis.
	The zigzag structure of the topmost S atoms is highlighted with dashed lines.
	(c)~A schematic phase diagram of \ce{NiS_{2-x}Se_{x}} reproduced from~\cite{yao_1996_prb}.
	PM, paramagnetic metal; AFI, antiferromagnetic insulator.
	The conditions investigated in the present manuscript are marked by black dots.
	(d)~A constant-current topographic image of \ce{NiS2}.
	(e)~A spatially averaged $\dv*{I}{V}$ spectrum at a 2D surface.
	The set-point bias voltage $V_\mathrm{s}=\SI[retain-explicit-plus]{+1}{\V}$, set-point current $I_\mathrm{s}=\SI{100}{\pA}$, and measurement temperature $T=\SI{77}{\K}$.
	}
	\label{fig:schematic}
\end{figure}

Transition-metal (TM) disulfides in the pyrite structure, in which TM$^{2+}$ and \ce{S2^{2-}} crystalize in the \ce{NaCl} structure with space group $Pa\bar{3}$~[Figs.~\ref{fig:schematic}(a) and (b)], offer a variety of electric and magnetic properties depending on its $d$-band filling~\cite{jarrett_1968_prl,bither_1968_ic, ogawa_1979_jap}.
\ce{FeS2} $(t_{2g}^6e_g^0)$ is a diamagnetic semiconductor, \ce{CoS2} $(t_{2g}^6e_g^1)$ is a ferromagnetic metal, \ce{NiS2} $(t_{2g}^6e_g^2)$ is an antiferromagnetic Mott insulator with half-filled $e_g$ bands, \ce{CuS2} $(t_{2g}^6e_g^3)$ is a superconductor, and \ce{ZnS2} $(t_{2g}^6e_g^4)$ is a semiconductor.

\ce{NiS2} undergoes a metal-to-insulator transition with bandwidth control, and it is considered to be a typical model as it does not change the crystal structure~\cite{imada_1998_rmp}.
The bandwidth control is achieved by Se substitution with S~\cite{yao_1996_prb,miyasaka_2000_jpsj,xu_2014_prl,han_2018_prb, jang_2021_nc} and an application of hydrostatic pressure~\cite{friedemann_2016_sr,hussain_2018_pb}.
The pristine \ce{NiS2} at an ambient pressure exhibits a noncolinear antiferromagnetic order below \SI{39}{\K}, and a weak ferromagnetic order below \SI{30}{\K} [Fig.~\ref{fig:schematic}(c)].

The resistivity of \ce{NiS2} tends to saturate at low temperatures even though it is an insulator.
It has been argued that the surface may be metallic and dominate the electric conduction at low temperatures~\cite{thio_1994_prb}.
The multi-channel conductance is supported by changing the crystalline grain size~\cite{rao_2011_jpc} and with heat treatments~\cite{clark_2016_jmmm}.
The neutron scattering measurements with susceptibility measurements suggest the surface contributes also to its magnetic state~\cite{thio_1995_prb}.
Additionally, a finite density of states (DOS) at the Fermi level by the photoemission spectroscopy~\cite{sarma_2003_prb, xu_2014_prl} further implies its metallic surface states.
More recent researches focusing on the surface anomalies convince the additional metallic channel~\cite{khatib_2021_prm} and imply magnetic anomalies at step edges~\cite{hartmann_2023_nanoscale, elkhatib_2023_prm}.

As surface conduction is implied in \ce{NiS2}, it is interesting to investigate the surface directly by using the spectroscopic-imaging scanning tunneling microscopy (STM).
An STM observation was reported for the metallic compounds \ce{NiS_{2-x}Se_{x}} $(x \geq 0.45)$~\cite{iwaya_2004_prb} whereas its insulating state has not been investigated.
In this manuscript, we report STM results for the Mott insulating side ($x=0$ and $0.45$).
In the three-dimensional (3D) crystal structure, we found that its two-dimensional (2D) surface is indeed insulating, contrary to the expectation from previous reports.
Interestingly, the Mott gap becomes smaller as approaches to one-dimensional (1D) step edges in the weak ferromagnetic insulator (WFI) and the paramagnetic
insulator (PI) phases.
Thus, we infer that the suggested conductance is happening at the 1D step edges rather than 2D surfaces.
Such edge effect was not observed in Se substituted metallic crystals \ce{NiS_{1.55}Se_{0.45}}.

\begin{figure}
	\includegraphics[width=\columnwidth]{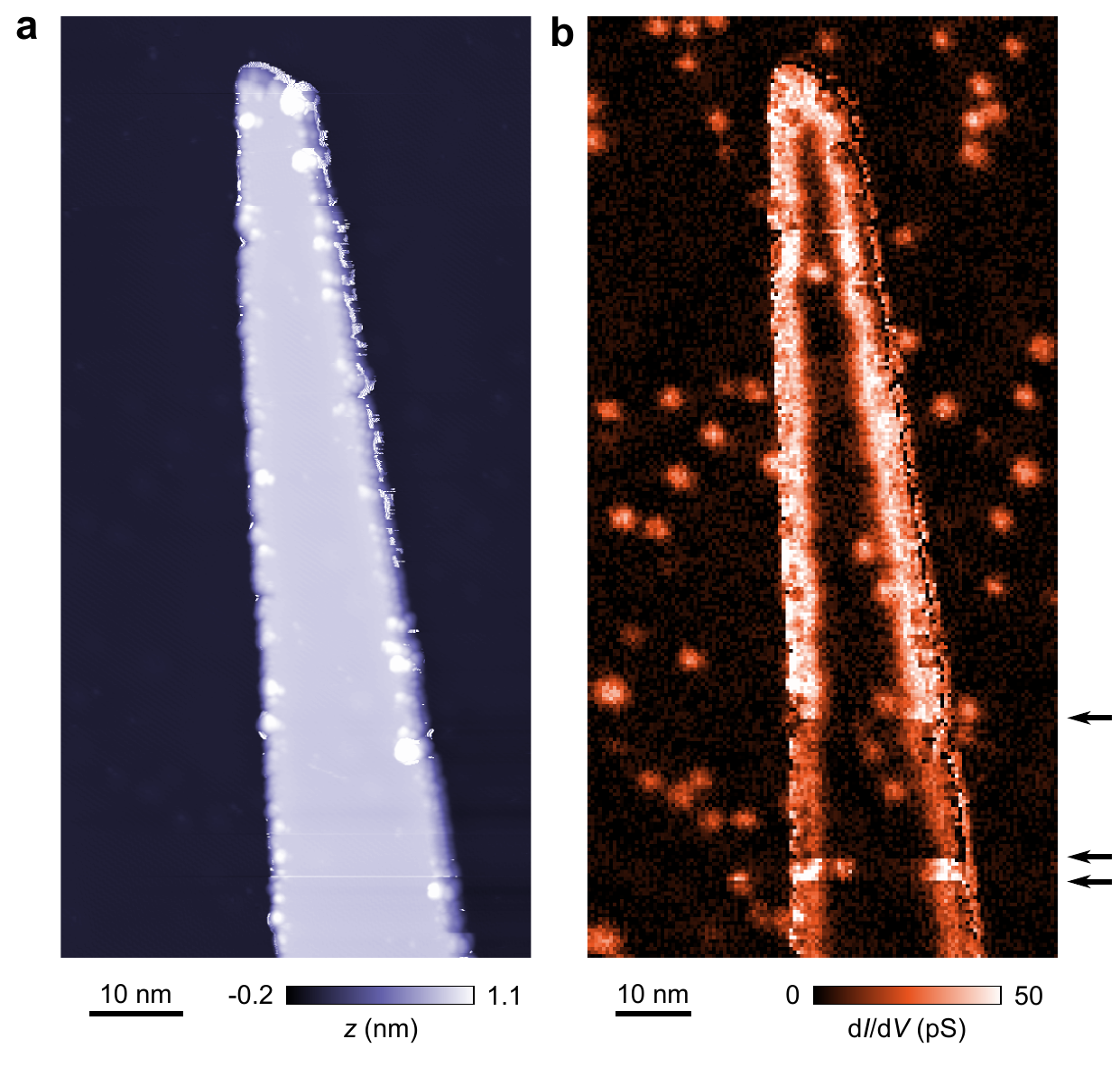}
	\caption{
		Edge states in \ce{NiS2} measured at \SI{77}{\K} (PI phase).
		(a)~A topographic image of a pseudo-island structure.
		(b)~A $\dv*{I}{V}$ map at \SI[retain-explicit-plus]{+200}{\mV} within the surface Mott gap taken in the same field of view as (a).
		Positions where the tip changed are marked with black arrows.
		$V_\mathrm{s}=\SI[retain-explicit-plus]{+0.9}{\V}$ and $I_\mathrm{s}=\SI{100}{\pA}$.
	}
	\label{fig:island}
\end{figure}

\begin{figure}
	\includegraphics[width=\columnwidth]{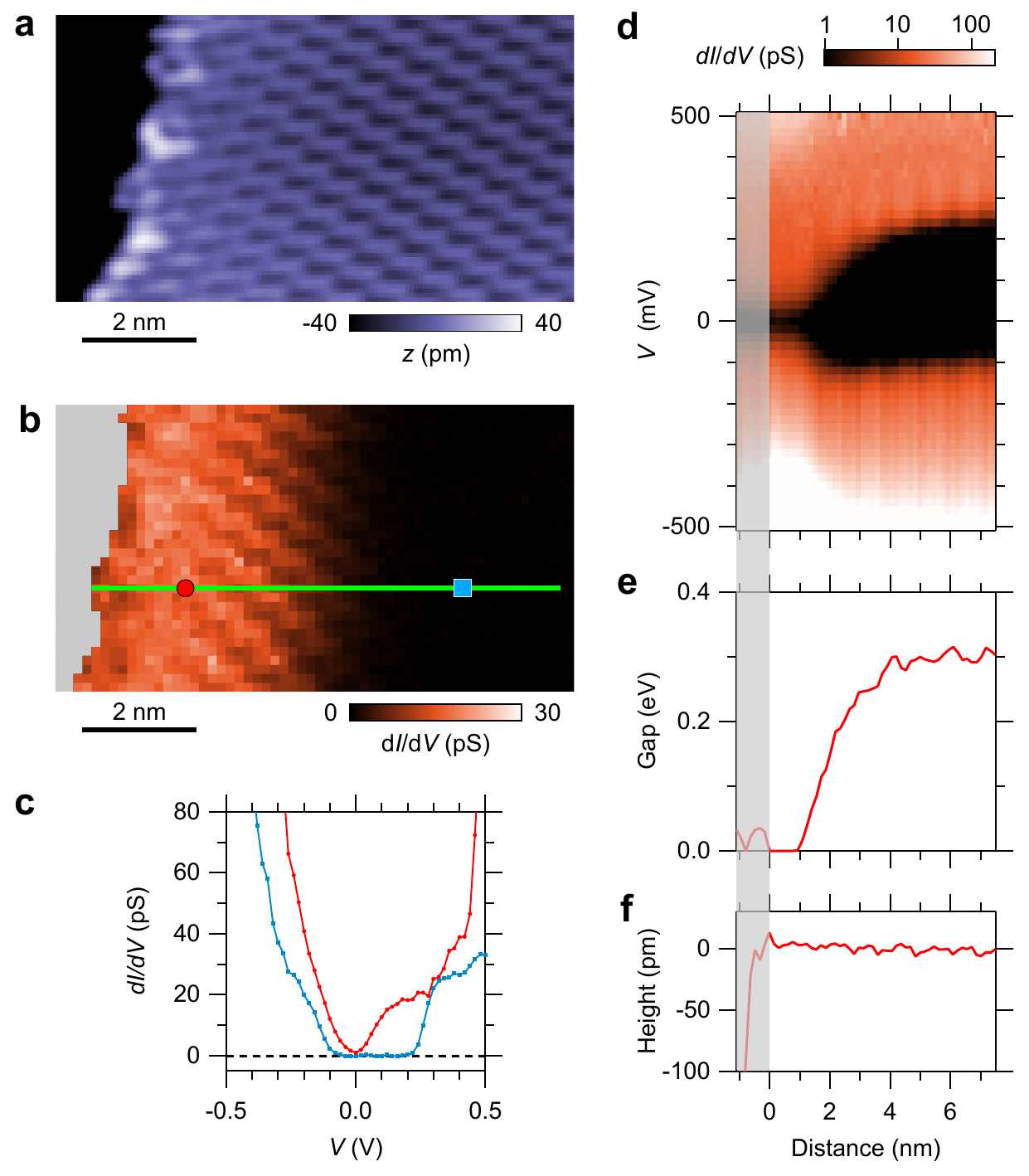}
	\caption{
		The edge anomaly in \ce{NiS2} measured at \SI{77}{\K} (PI phase).
		(a)~A topographic image near a step edge.
		(b)~A $\dv*{I}{V}$ map at \SI[retain-explicit-plus]{+200}{\mV} within the surface Mott gap taken in the same field of view as (a).
		Points at $z<\SI{-50}{\pm}$ are not shown.
		(c)~$\dv*{I}{V}$ spectra near (red circle) and away (blue square) from the step edge.
		(d)~Tunneling conductance spectra along the green line in (b).
		(e)~The variation of the insulating gap defined at \SI{1}{\pico \siemens}.
		(f)~The line profile of the simultaneously taken topographic image.
		$V_\mathrm{s}=\SI[retain-explicit-plus]{+1.0}{\V}$ and $I_\mathrm{s}=\SI{100}{\pA}$.
	}
	\label{fig:pi}
\end{figure}

\begin{figure}
	\includegraphics[width=\columnwidth]{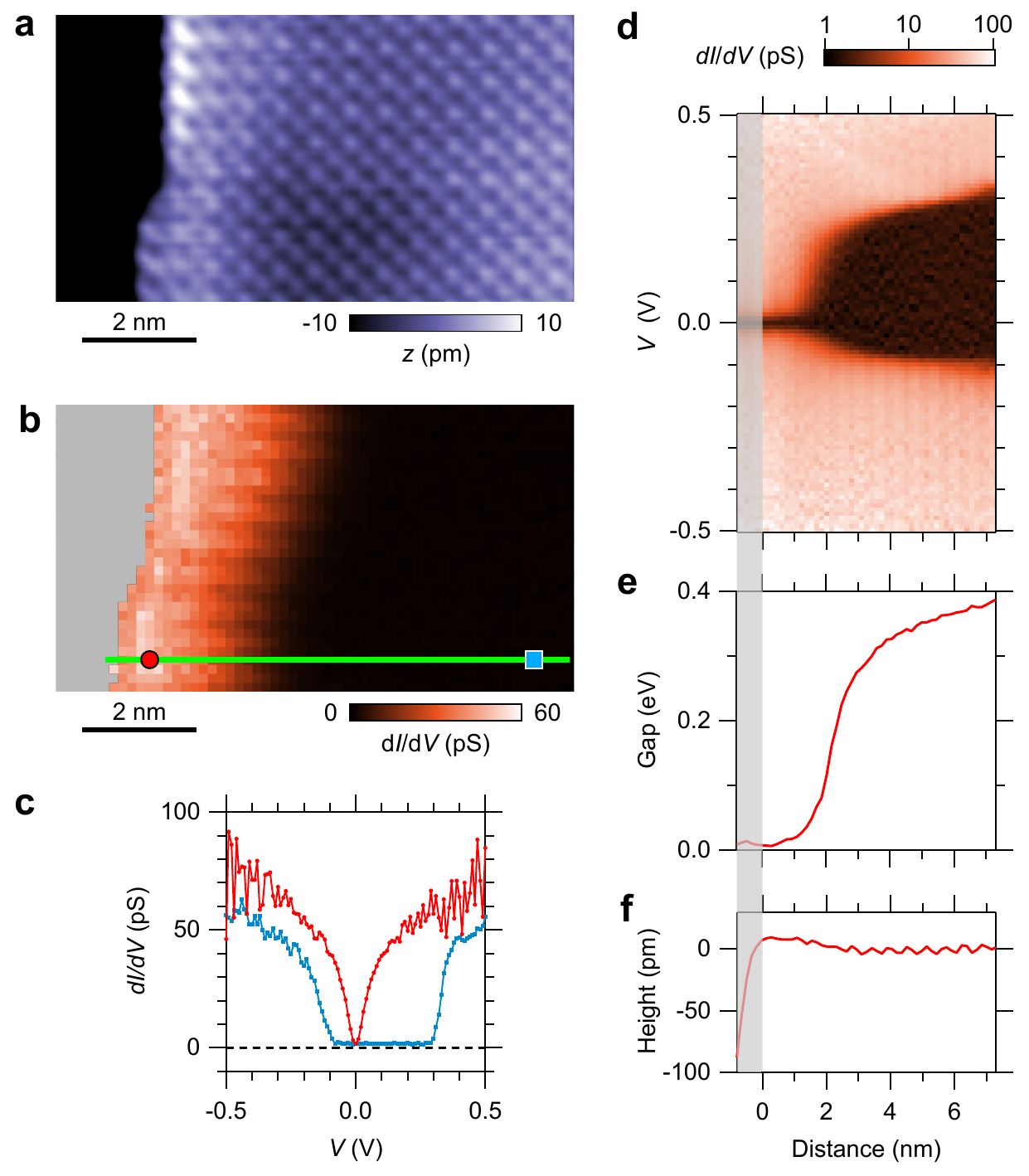}
	\caption{
		The edge anomaly in \ce{NiS2} measured at \SI{5}{\K} (WFI phase).
		(a)~A topographic image near a step edge.
		The image is Fourier-filtered to suppress high-frequency noise.
		(b)~A $\dv*{I}{V}$ map at \SI[retain-explicit-plus]{+200}{\mV} within the surface Mott gap in the same field of view as (a).
		Points at $z<\SI{-50}{\pm}$ are not shown.
		(c)~$\dv*{I}{V}$ spectra near (red circle) and away from the step (blue square).
		(d)~Tunneling conductance spectra along the green line in (b).
		(e)~The variation of the insulating gap defined at \SI{2.5}{\pico \siemens}.
		(f)~The line profile of the simultaneously taken topographic image.
		$V_\mathrm{s}=\SI[retain-explicit-plus]{+1.5}{\V}$ and $I_\mathrm{s}=\SI{200}{\pA}$.
	}
	\label{fig:wfi}
\end{figure}

\begin{figure}
	\includegraphics[width=\columnwidth]{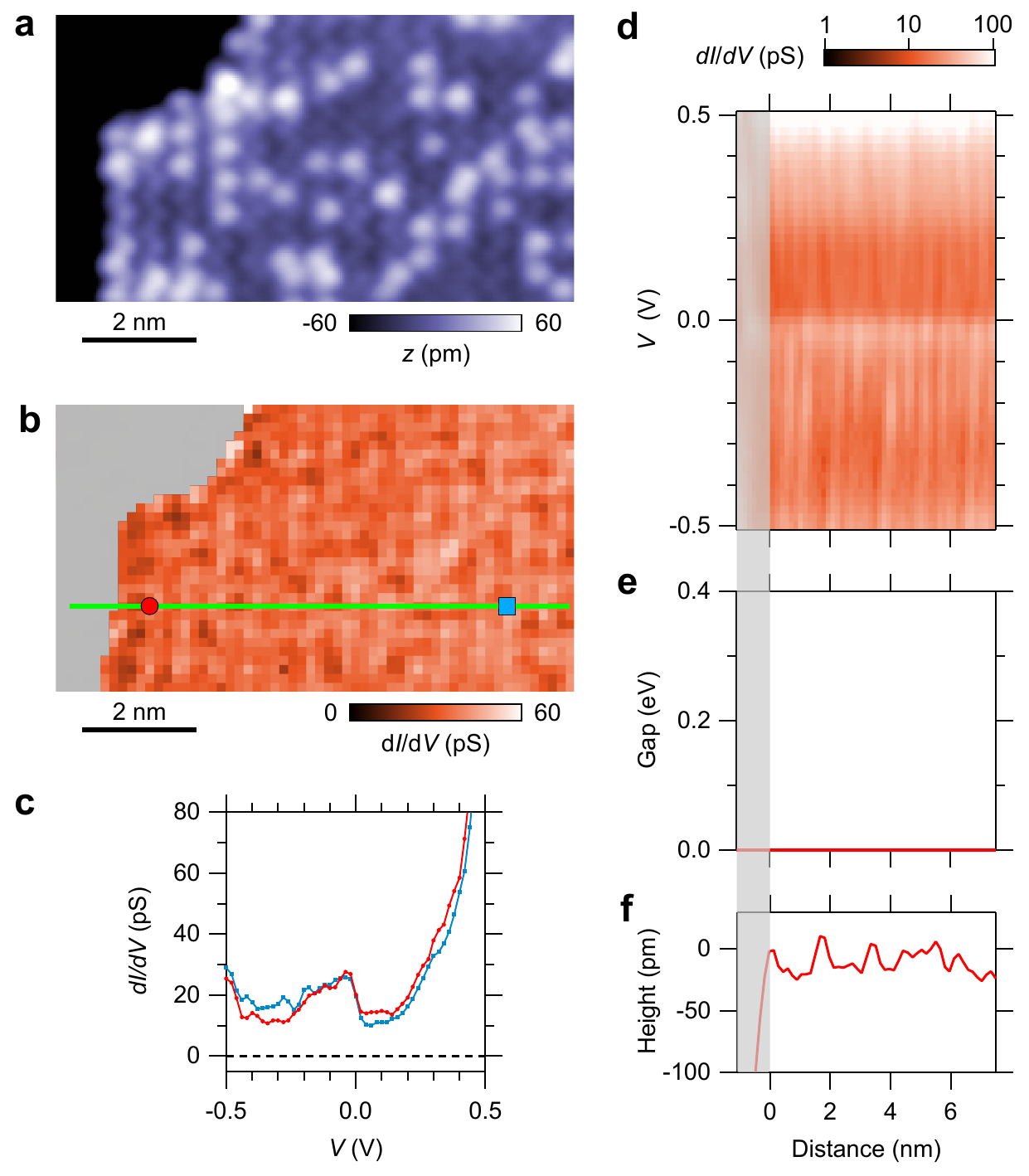}
	\caption{
	Absence of a noticeable edge effect at the boundary of the PI and the AFM phases in \ce{NiS_{1.55}Se_{0.45}} measured at \SI{77}{\K}.
	(a)~A topographic image near a step edge.
	Se atoms appear as bright protrusions while S atoms appear darker in the same zigzag structure.
	(b)~A $\dv*{I}{V}$ map at \SI[retain-explicit-plus]{+200}{\mV} in the same field of view as (a).
	Points at $z<-50$~pm are not shown.
	(c)~$\dv*{I}{V}$ spectra near (red circle) and away from the step (blue square).
	(d)~Tunneling conductance spectra along the green line in (b).
	(e)~The variation of the insulating gap defined at \SI{2.5}{\pico \siemens}.
	(f)~The line profile of the simultaneously taken topographic image.
	$V_\mathrm{s}=\SI[retain-explicit-plus]{+1.0}{\V}$ and $I_\mathrm{s}=\SI{100}{\pA}$.
	}
	\label{fig:pi_afm}
\end{figure}

\ce{NiS_{2-x}Se_{x}} $(x=0 \, \mathrm{and} \, 0.45)$ single crystals were grown with the chemical vapor transport method~\cite{miyasaka_2000_jpsj}.
For STM measurements at \SI{77}{\K},
the samples were cleaved in an ultra-high vacuum chamber (\SI{\sim e-8}{\Pa}) at around \SI{77}{\K} to expose clean and flat \hkl(0 0 1) surfaces and then transferred to a microscope without breaking vacuum.
As scanning tips, tungsten wires were used after electro-chemical etching in
\ce{KOH} aqueous solution, followed by tuning using field ion microscopy and
controlled indentation at clean Cu\hkl(1 1 1) surfaces.
Measurements at \SI{5}{\K} were performed with the Omicron low-temperature scanning tunneling microscopy (LT-STM) system.
The crystals were cleaved in the ultra-high vacuum chamber at room temperature.
A tungsten tip attached to a qPlus sensor was used after controlled indentation at clean Au\hkl(1 1 1) surfaces.
Tunneling-conductance $\dv*{I}{V}$ curves were obtained with numerical differentials of tunneling current $I$ versus bias voltage $V$ curves.

An STM topographic image of the cleaved surface is shown in Fig.~\ref{fig:schematic}(d).
One type of surface is available in this material because the S-S dimer bonds may not be broken.
The zigzag structure does not contradict the model structure as in Fig.~\ref{fig:schematic}(b).
The observed lattice constant is around 560~pm, which agrees with the literature value of \ce{NiS2}~\cite{kwizera_1980_prb}.

Tunneling conductance curves were measured on such cleaved surfaces at \SI{77}{\K}, in the PI phase~[Fig.~\ref{fig:schematic}(e)].
The observed $\dv*{I}{V}$ spectra are approximately proportional to the local density of states (LDOS), and hence we can directory refer to the insulating gap.
The measured insulating gap width is about \SI{0.3}{\eV} at \SI{77}{\K}, which is comparable to the reported gap width of the bulk~\cite{kautz_1972_prb}.
Contrary to the speculations of the previous reports~\cite{thio_1994_prb, rao_2011_jpc, clark_2016_jmmm, thio_1995_prb, sarma_2003_prb, xu_2014_prl, khatib_2021_prm}, the 2D surfaces are found to be insulating.

As the surface of \ce{NiS2} is found to be insulating, metallic states may originate from some defects or from some specific structures.
We investigated the electronic states near step edges because the lack of a good cleavage plane in \ce{NiS2} results in a large number of step-terrace structures at cleavages.
The electronic state just below step edges cannot be investigated with STM since a finite radius of a scanning tip causes tunneling current through the side of the tip.
Such artifacts of the tip shape are avoided on terraces above steps.
Thus, we focus on the electronic states above the steps in the following, and the data are considered only for the topographic height $\qty|z| < \SI{50}{\pm}$.

A step-terrace structure with 2-unit-cell hight~[Fig.~\ref{fig:island}(a)] was investigated with the spectroscopic-imaging technique, where the $\dv*{I}{V}$ conductance curves are measured at each pixel of the image.
Figure~\ref{fig:island}(b) shows a $\dv*{I}{V}$ map at \SI[retain-explicit-plus]{+200}{\mV}, representing the local DOS distribution within the insulating gap.
The in-gap states appear all around the step edges.
Thus, the behavior is not constrained to a specific orientation of the step with respect to the crystalline axis.
Although the tip condition changed several times during the measurements as marked with black arrows, the in-gap state is observed irrespective of the tip condition.

A magnified view shows a more detailed structure of the in-gap states.
The step was obtained from the same crystal at a different cleavage, and the step height corresponds to 1 unit cell.
The topographic image is shown in Fig.~\ref{fig:pi}(a), and a DOS map at \SI[retain-explicit-plus]{+200}{\mV} representing the in-gap edge states is shown in Fig.~\ref{fig:pi}(b).
These images are deformed due to the thermal drift but the atomic registry is maintained.
Figure~\ref{fig:pi}(c) compares the tunneling spectra near and away from the step edge.
The insulating gap is almost closed near the step edge, resulting in a quasimetallic state in the 1D step edge.
The spatial distribution of the tunneling spectra near the step edge is shown in Figs.~\ref{fig:pi}(d) -- (f).
The in-gap states near the step edge are not restricted to the step edge, but they spread over several nanometers into the terrace.

Some bright spots in $\dv*{I}{V}$ are observed on the terraces [Fig.~\ref{fig:island}(b)].
These spots probably originated from impurities embedded under the surface.
Although the $\dv*{I}{V}$ spectra on these impurities are very similar to those near the step edges, it is unlikely that the observed gap collapse is associated with the impurities.
This is because the in-gap states homogeneously distribute around step edges regardless of the position of impurities.
Therefore, the metallic state is intrinsic to step edges of \ce{NiS2}.

The location of the gap with respect to the chemical potential depends on samples (compare Fig.~\ref{fig:schematic}(e) and Fig.~\ref{fig:pi}(c)).
They probably originated from some doping effects from impurities.
We confirmed that the gap closes toward the Fermi energy near step edges wherever the surface gap is.

To investigate the effect of magnetic order on the in-gap state near the step edge, we performed spectroscopic-imaging measurements near step edges also in the WFI phase at \SI{5}{\K}.
A \ce{NiS2} crystal from a different batch was used.
Atomically sharp step edges are resolved in the topographic image in Fig.~\ref{fig:wfi}(a), where the tip probably feels the symmetry of Ni atoms as well.
The step height is \SI{4}{\nm}, and the zigzag structure of S atoms is perpendicular to the step.
As shown in Figs.~\ref{fig:wfi}(b) -- (f), the insulating gap gradually collapses over a few nanometers as approaching the step edge, similar to the paramagnetic phase.

A natural question is whether the edge-induced effect is observed when the insulating gap is suppressed by the bandwidth control.
To this end, we investigated \ce{NiS_{2-x}Se_x} (nominal $x=0.45$), which is located around the boundary of the PI and the antiferromagnetic metal (AFM) phases [Fig.~\ref{fig:schematic}(c)].
As shown in Fig.~\ref{fig:pi_afm}, the topographic image exhibits two atomic sites with different heights.
Se atoms are considered to appear higher in constant-current topographic images~\cite{iwaya_2004_prb}.
From the ratio between the numbers of higher and lower atoms on the surface, $x$ is estimated to be 0.43, which agrees with the nominal amount of Se.
A $\dv*{I}{V}$ map is measured at a step edge with 0.5-unit-cell hight [Figs.~\ref{fig:pi_afm}(a) and (b)].
There is no gap in the tunneling spectra as shown in Fig.~\ref{fig:pi_afm}(c).
A line profile clarifies that any edge effect was not evident at any bias voltage within \SI{\pm 0.5}{\V} [Figs.~\ref{fig:pi_afm}(d) -- (f)].

The surface conduction in \ce{NiS2} has been proposed in the previous reports~\cite{thio_1994_prb,thio_1995_prb,rao_2011_jpc,clark_2016_jmmm,khatib_2021_prm}.
The present observations clarify that the 2D surfaces are as insulating as their bulk, and in fact, their 1D edges are more metallic.
Since the in-gap states near step edges can provide an additional channel to the conductance, whose amount is directly related to the surface area, we consider that the previous observations are consistent for 1D-edge conductance.
The 1D-edge channel can also infer why the previous angle-resolved photoemission spectroscopy measurements did not report dispersive features within the Mott gap even though a finite DOS at the Fermi level is observed~\cite{matsuura_1998_prb,sarma_2003_prb,xu_2014_prl,jang_2021_nc}.

The question is what causes the observed metallic edge state. One possible scenario concerns a bandwidth-controlled Mott insulator-to-metal transition, which occurs when the onsite Coulomb repulsion $U$ becomes smaller with respect to the bandwidth $W$.
In bulk \ce{NiS2}, physical or chemical pressures induce the bandwidth-controlled Mott insulator-to-metal transition. However, $U$ and $W$ are expected to become larger and smaller, respectively, at the step edges due to weaker screening and hopping, which contradict the observed metallic edge state.
Particular edge termination may effectively cause local doping, resulting in a metallic edge, as in the case of the metallic surface of \ce{V_{2-x}Cr_xO_3} with excess vanadyl cations~\cite{lantz_2015_prl}. In the present case, however, the metallic edge states appear irrespective of the edge direction, ruling out the possibility of a termination-dependent metallic state.
Another possible cause of the metallic state is topological edge states~\cite{tang_2017_np, schindler_2018_np, xu_2020_nature, shumiya_2022_nm, yin_2022_prl}. Although the topological aspect of the band structure has theoretically been argued in \ce{NiS2} for $U = 0$~\cite{xu_2020_nature}, we are not aware of such an expectation for a more realistic situation with $U \ne 0$.
More experimental and theoretical work is indispensable to clarify the underlying mechanism of the 1D metallic edge states in pyrite \ce{NiS2}.

In summary, the spectroscopic-imaging STM measurements on the charge-transfer type Mott insulator \ce{NiS2} clarified that the 2D surface is as gapped as the 3D bulk.
The Mott gap collapses near the step edges in the WFI and the PI phases, exhibiting a quasimetallic DOS.
This behavior was not observed at the phase boundary between the PI and the AFM phases.
The spatially resolved results obtained in this work indicate that the surface conduction so far inferred from macroscopic measurements~\cite{thio_1994_prb, rao_2011_jpc, clark_2016_jmmm, khatib_2021_prm} are not related to the surface but associated with the step edges.
Different types of ``surface'' effects have been found in various $3d$ transition-metal disulfides~\cite{bronold_1994_ss, herbert_2013_ss, limpinsel_2014_ees, walter_2017_prm, walter_2020_sa, wu_2009_jpcm, schroter_2020_sa}.
We anticipate that their systematic spatially-resolved investigations may help to understand the role of step edges in the pyrite structure.

\section*{Acknowledgement}
The authors acknowledge T. Yoshida, Y. Kohsaka, T. Machida, and C. J. Butler for discussion.
This work was supported by Grant-in-Aid JSPS KAKENHI Grant Nos. JP19H01855, JP21H04652, JP21H05236, JP21K18181, JP22H04496, and JP22H05448.

\bibliography{reference.bib}

\end{document}